\documentclass{iopart}

\usepackage{epsfig}

\begin{document}

\jl{6}   
\newcommand{\eqn}[1]{(\ref{eqn:#1})}
\newcommand{\fig}[2]{\ref{fig:#1}{\sl #2}}

\title[]{On the convergence of Regge calculus to general relativity}

\author{Leo C. Brewin and Adrian P. Gentle\footnote[1]{Permanent address:
    Theoretical Division (T-6, MS B288), Los Alamos National
    Laboratory, Los Alamos, NM 87545, USA.}}
\address{Department of Mathematics and Statistics,\\ Monash University, 
  PO Box 28M, Victoria 3800, Australia.}

\begin{abstract}
  Motivated by a recent study casting doubt on the correspondence
  between Regge calculus and general relativity in the continuum
  limit, we explore a mechanism by which the simplicial solutions can
  converge whilst the residual of the Regge equations evaluated on the
  continuum solutions does not.  By directly constructing simplicial
  solutions for the Kasner cosmology we show that the oscillatory
  behaviour of the discrepancy between the Einstein and Regge
  solutions reconciles the apparent conflict between the results of
  Brewin and those of previous studies.  We conclude that solutions of
  Regge calculus are, in general, expected to be second order accurate
  approximations to the corresponding continuum solutions.
\end{abstract}
\pacs{04.20.-q, 04.25.Dm}

\setcounter{footnote}{0}


\section{Introduction}

Regge calculus \cite{regge61} is a discrete theory of gravity which
replaces the smoothly curved spacetime of general relativity with a
lattice.  The curvature of the lattice spacetime is concentrated
entirely on the two-dimensional hinges of the four-dimensional lattice
cells.

Regge calculus holds much promise for the numerical investigation of
both classical and quantum gravity. Although the lattice approach
appears well suited to numerical applications, progress in the field
has been slow.  Only recently have the first completely generic
four-dimensional numerical simulations been performed
\cite{gentle98,gentle99}.

After the recent papers by Brewin \cite{brewin95} and M.~Miller
\cite{miller95}, and despite the proven track record of the Regge
calculus, a lively debate arose \cite{private} as to whether or not
solutions of the Regge equations would converge to solutions of the
Einstein equations in some suitable limit.  Neither Brewin or
M.~Miller directly computed solutions of the Regge equations.
Instead, they took the somewhat easier approach of evaluating the
Regge equations on an exact solution of Einstein's equations. They did
this using sophisticated interpolation schemes, based on geodesics, to
map a range of Einstein solutions onto simplicial lattices. The
residual of the resultant Regge equations, calculated using these
interpolated lattice edge lengths, was then examined in the limit of
very fine lattice discretisations. Brewin observed that the residual
scaled as ${\cal O}(1)$ as the lattice was refined, and he inferred
that solutions of the Regge equations, in generic spacetimes, would
not converge to solutions of the Einstein equations in the limit of
fine discretisations. M.~Miller chose a slightly different measure for
the residual of the Regge equations, but when expressed in terms of
Brewin's measure, M.~Miller's results are consistent with those
obtained by Brewin. Despite this M.~Miller does draw sharply different
conclusions, based upon further calculations associated with averages
of the Regge equations, namely that solutions of the standard Regge
equations will converge as the fourth order in the lattice spacing to
solutions of the Einstein equations.  Similar behaviour, in which the
residual converges only over averages of the equations,  has been
observed by Sorkin \cite{sorkin75} in the context of massless scalar
fields on a 2-dimensional simplicial space.

The situation became even more confused after the work of Gentle and
W.~Miller \cite{gentle98}, who demonstrated explicit quadratic
convergence of particular solutions of Regge calculus to a solution,
the Kasner spacetime, of general relativity. This spacetime was one of
the test cases used by Brewin and M.~Miller. This seems most odd -- we
appear to have a convergent set of solutions from an apparently
non-convergent set of equations. How can this be?

In this paper we explore a possible explanation for this behaviour, as
proposed by Brewin \cite{brewin95}, which is consistent with all
previous numerical studies.  It is important to note that the direct
solution of the Regge equations has always generated solutions which
converge to the corresponding solutions of general relativity.
Although the observed rate of convergence is dependent upon the
particular lattice and symmetry restrictions imposed, when the lattice
construction allows the full expression of the gravitational degrees
of freedom, second order convergence of the lattice solutions to the
continuum has always been observed \cite{gentle99}.

We begin with a more complete description of the problem in section
\ref{sec:details}, and then discuss a possible explanation in section
\ref{sec:explain}.  Finally, in section \ref{sec:numerics}, we present
numerical evidence to support our proposal.

\section{Investigating the convergence of Regge calculus}
\label{sec:details}

There are various ways in which the convergence properties of a
numerical technique can be explored.  The most rigorous approach
involves the direct comparison of the approximate equations and the
full system, concentrating on the leading order discrepancies.  The
underlying assumption is that the solutions of the approximate
equations will converge at the same rate as the approximate equations
themselves converge to the original system.

The Regge and Einstein equations are too complex for such a direct
approach to be beneficial.  Not only are the equations inherently
complex non-linear systems, but before such an analysis can proceed
one must choose a lattice upon which to express the Regge equations,
together with a coordinate system in which to write out the Einstein
equations.  Once these decisions have been made, we are faced with a
more problematic choice: how are the two sets of equations to be
compared?  In general there are more Regge equations (one per lattice
edge) than Einstein equations (ten per spacetime event).

There is no clear prescription for averaging the lattice equations to
obtain the correct number of Einstein equations in the continuum
limit.  Brewin \cite{brewin95} and M.~Miller \cite{miller95} have both
developed their own schemes for averaging the Regge equations. A
direct assault on the convergence properties of Regge calculus is
therefore likely to be applicable only on a specific lattice, with a
particular choice of averaging in the continuum limit.  That is, the
results would indicate that, in some averaged sense, the Regge
equations converge to the Einstein equations. However, the result
would not constitute a proof valid for all possible lattice choices.

Given these limitations, it is natural to ask if similar results can
be obtained by directly solving the Regge equations and comparing the
lattice solutions to corresponding Einstein spacetimes.  Brewin and
M.~Miller have investigated the inverse problem: given a known
(analytic) solution of general relativity, and interpolating that
solution onto a lattice, what can be said about the convergence of the
Regge equations?

The approach taken by Brewin and M.~Miller was to introduce some
discretisation process in the smooth manifold on which the Einstein
equations are defined.  This provides a way to map the smooth metric
solutions to a lattice, giving a set of lattice edge lengths
$\mathbf{L}_E$ derived from the continuum metric.  Both authors used
geodesic lengths to map continuum information onto the lattice
spacetime.  These new ``continuum'' lattice edge lengths do not in
general satisfy the Regge equations, but we can evaluate the residual
\begin{equation}
  \label{eqn:residual}
  r = | \mathbf{R}(\mathbf{L}_E)  |
\end{equation}
which is an indication of how well the interpolated Einstein solution
satisfies the Regge equations.  It is this residual, using an
appropriately chosen norm, which both Brewin and M.~Miller considered
\cite{brewin95,miller95}.

Brewin observed that the residual of the simplicial equations remained
roughly constant as the lattice was refined on a fixed region of
spacetime. This result led Brewin to question the validity of Regge
calculus, since any useful numerical scheme must converge to the
underlying solution of the partial differential equations as the
resolution is improved.  M.~Miller observed second order convergence
of the residual, for any smooth metric, whether or not they were
solutions of Einstein's equations. Had he used the same norm for the
residual as used by Brewin we believe he too would have observed
${\cal O}(1)$ convergence of the residuals.

It is important to note that the experiments of Brewin and M.~Miller
do not directly evaluate the convergence of the numerical solutions.
Rather, they investigate the convergence of the lattice equations to
the Einstein equations.  As the lattice is refined, it is reasonable
to expect that the interpolated Einstein solution will satisfy the
Regge equations increasingly accurately.  This, however, is not what
Brewin observed numerically.

\section{A possible explanation}
\label{sec:explain}

The observations of Brewin are particularly puzzling in light of the
many previous applications of Regge calculus.  Almost every numerical
application of the method has displayed convergence towards the
corresponding continuum solution, with most studies indicating that
numerical Regge calculus is a second order accurate approximation to
general relativity.

The edge lengths measured in the Regge lattice ($L$) and the
corresponding interpolated Einstein edges (${L}_E$, obtained by
assigning geodesic lengths calculated in the continuum) are related as
\begin{equation}
  \label{eqn:error}
  L = L_E + {\cal O}(\delta^{p+1})   
\end{equation}
with previous numerical studies suggesting that $p = 2$ (see Gentle
\cite{gentle99} for case studies and a general review).  Throughout
this paper we assume that $\delta$ is a typical length scale in the
lattice, and noting that the edges themselves are of this magnitude,
we say that edges which satisfy equation \eqn{error} are $p$-order
accurate approximations to the continuum solution.

Brewin \cite{brewin95} has proposed a mechanism whereby the solutions
of the Regge equations converge to the corresponding Einstein
solutions, while at the same time the residual of the Regge equations
evaluated on the interpolated Einstein lattice edges does not
converge. The key to Brewin's proposal is allowing the functional form
of the error terms (the discrepancy between the Einstein and Regge
solutions) to depend on the discretisation scale $\delta$.  This can
be clearly seen from a toy model.

Suppose ${\cal L}$ is a second order differential operator, and $y(x)$ is
a solution of the equation
\begin{equation}
  {\cal L}y = 0.
\end{equation}
Brewin considers the function
\begin{equation}
  \tilde y_\delta(x) =  y(x) + \delta^2 f(x/\delta)
\end{equation}
for some arbitrary scalar $\delta$ and an arbitrary, though bounded,
function $f(x)$.  Considering the difference
between the solutions we find that 
\begin{equation}
  | y(x) - \tilde y_\delta(x) | = {\cal O}(\delta^2),
\end{equation}
indicating that the solutions differ only  by ``second order'' terms.
Noting that $\tilde y_\delta(x)$ is a solution of some other
related equation  $\tilde {\cal L}_\delta$, 
\begin{equation}
  \tilde {\cal L}_\delta \tilde y_\delta = 0,
\end{equation}
where $\tilde{\cal L}_\delta$ is also a second 
order differential operator (though different from $\cal L$),
we find that the ``residual'' of $\tilde y_\delta(x)$ with respect
to the original operator is 
\begin{equation}
 | {\cal L}\,  \tilde y_\delta | = {\cal O}(1).
\end{equation}
This toy model embodies precisely the properties observed by Brewin --
second order convergence of the solutions \cite{gentle99}, with no
corresponding convergence observed in the Regge equations when they
are evaluated on exact solutions interpolated from the continuum
\cite{brewin95}.  The discrepancy between the two solutions in this
toy model is seen to be a wave-like disturbance with frequency
proportional to $1/\delta$.

In the continuum limit it is reasonable to expect that the discrete
Regge equations (or a weighted average over them) approach a system of
differential equations.  Moreover, consideration of the Einstein
equations leads us to expect that the limiting form of the Regge
equations is a set of second order non-linear equations.  Furthermore,
previous numerical experiments suggest that Regge calculus is a second
order method; we expect $p=2$.  The toy model considered above leads
us to expect that if the lattice solutions differ from the continuum
solutions by terms with frequencies proportional to $1/\delta$, it is
not unreasonable that the residuals of the Regge equations remain
roughly constant as the resolution is improved.

This is precisely the behaviour observed by Brewin \cite{brewin95};
the solutions converge even though the residual of the equations do
not.  This explanation relies on the existence of high frequency, low
amplitude waves in the simplicial solutions; a possibility not ruled
out by any of the largely low-resolution applications of Regge
calculus to date.  In fact, one study hints at the existence of
precisely this type of wave-like structure in the Regge solutions
\cite{gentle98}.  In the next section we construct high resolution
solutions of the Regge equations in order to gain insight into the
fine-scale behaviour of the lattice solutions.

\section{Numerical solution of the Regge equations}
\label{sec:numerics}

In this section we solve the Regge equations for the vacuum Kasner
cosmology using a $(3+1)$-dimensional formulation of Regge calculus
described elsewhere \cite{gentle98}.  The initial value problem is
solved and the lattice is evolved subject to simplicial lapse and
shift conditions.

The Kasner cosmology is a homogeneous anisotropic exact solution of
the vacuum Einstein equations, and is the prototype for generic
velocity dominated cosmological singularities.  It also provides a
model for epochs in the mixmaster cosmology between bounces.  The
metric may be written in the form
\begin{equation}
  ds ^2 = -dt^2 + t^{p_1} dx^2 + t^{p_2} dy^2 + t^{p_3} dz^2,
\end{equation}
where the field equations reduce to a pair of algebraic constraints,
\begin{equation}
  p_1 + p_2 + p_3 = p_1^2 + p_2^2 + p_3^2 = 1.
\end{equation}
The Kasner exponents $(p_1,p_2,p_3)$ are a one-parameter set,
and the solutions include the flat-spacetime case $(0,0,1)$.

The previous application of Regge calculus to the Kasner cosmology was
viewed as a benchmark of the numerical technique, and as such the
convergence of the solutions was an important issue.  It was found
that the lattice solution converged to the continuum as the second
power of the typical lattice spacing \cite{gentle98}. This earlier
study used a fixed lattice resolution, containing 128 vertices
arranged in two offset cubic grids each consisting of
$4\times4\times4$ vertices. Reducing the overall scale of the lattice,
combined with the $T^3$ topology of the Kasner solution, allowed the
convergence rate to be estimated.

In this paper we directly subdivide the lattice whilst fixing the size
of the spatial region. This is a more general and demanding
convergence test, as any long wavelength modes excluded in the
previous study are now resolvable.  Since the major issue is
convergence we refine the lattice in only one dimension, but reduce
the scale of the remaining spatial axes to avoid the creation of long
skinny triangles.  Such long and skinny triangular elements are known
to cause instabilities in finite element calculations.

This effectively ``one-dimensional'' model allows convergence analysis
to be conducted whilst containing the computational scale of the
problem.  All calculations are performed on a lattice containing two
offset cubic grids which are fully subdivided into triangles,
tetrahedra and four-simplices.  Each cubic grid contains $2^n\times
4\times4$ vertices ($N \equiv 2^n=4,8,\ldots,1024$), with the $x$-axis
having a fixed length of $X=10$, and the remaining axes scaled to
compensate; $Y=Z=40/2^n$.  This ensures that the spatial edges
surrounding a vertex are all of the same order of magnitude.  The
typical spatial resolution in each model is thus $\delta\approx
10/2^n$.

The full four-dimensional two-slice initial value problem is solved at
$t=1$, after which the lattice is evolved to $t=2$ using the
$(3+1)$-dimensional Sorkin evolution scheme and ``geodesic slicing''
conditions; that is, with unit lapse and zero shift.  The time step is
chosen for each resolution to give a Courant factor of $dt/l \approx
0.2$; for $N=512$ we chose $dt = 3.9\times 10^{-3}$.  However, the
main results of this paper were found to be largely insensitive to the
choice of timestep, provided it satisfied the Courant condition.  For
full details of the evolution and initial value algorithms, see Gentle
and W.~Miller \cite{gentle98}.  The results of the Regge evolution are
compared with the continuum solution by performing a least-squares fit
of the function
\begin{equation}
  \label{eqn:fit}
  L(t) = L_0 t^{p_r}
\end{equation}
to the time evolution of each class of the axis-aligned spatial edges,
$L_x, L_y$ and $L_z$.  In agreement with previous studies, the lattice
solution is found to remain spatially homogeneous throughout the
evolution to within roundoff error.  The typical standard deviation of
the edges also remained below the limit of numerical accuracy.

\begin{figure}[e]
  \begin{center}
    \epsfig{file=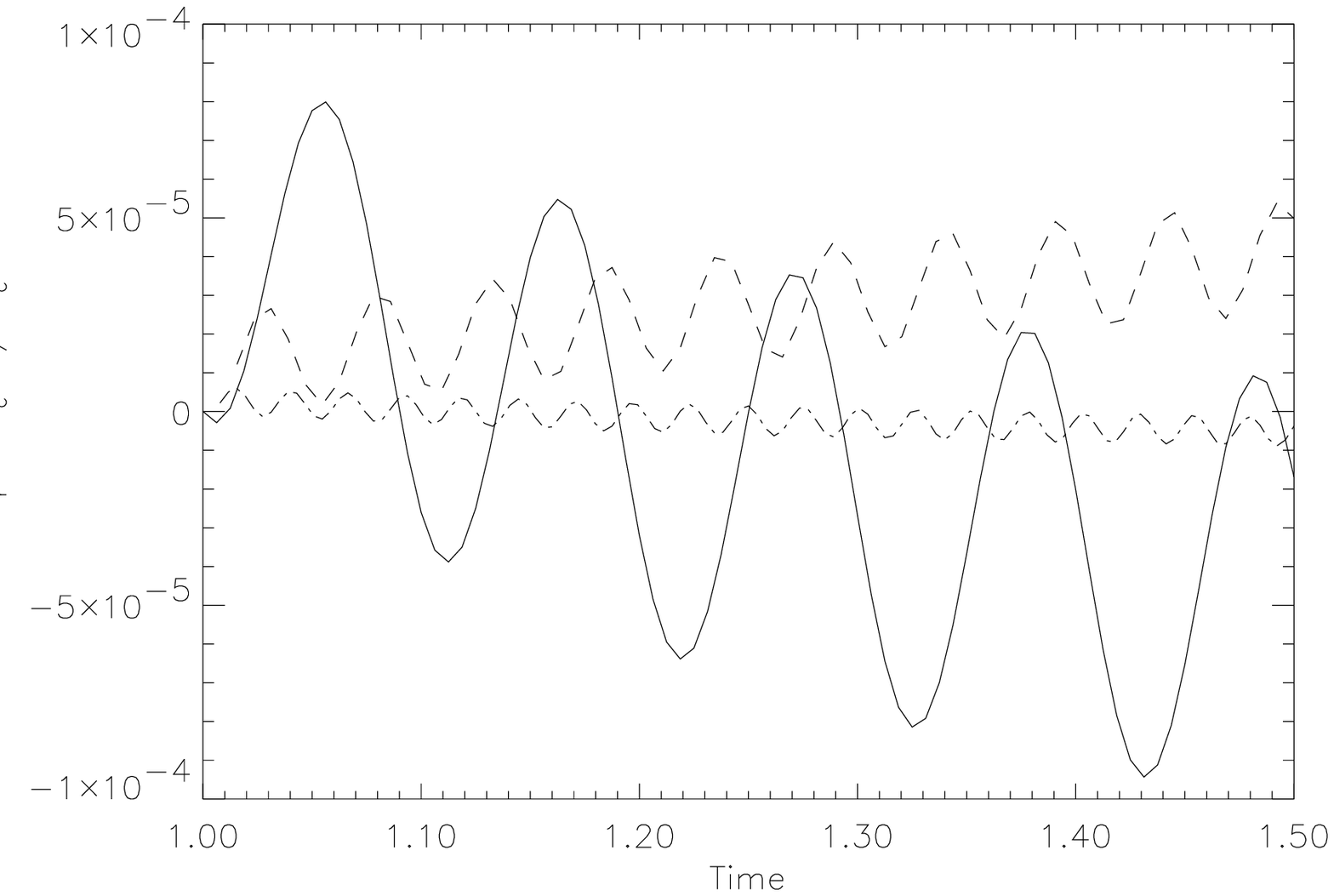,width=4.0in}
  \end{center}
  \caption{The fractional error $\delta L/L$ of a typical edge in the
    Regge lattice as a function of time.  Results are shown for
    spatial edges aligned with the $x$-axis, using lattice resolutions
    of $128\times 4\times 4$ ($\full$), $256\times 4\times 4$
    ($\broken$), and $512\times 4\times 4$ ($\chain$) vertices.  In
    all cases the total length of the lattice along the $x$-axis is
    $10.0$ units. The magnitude of the observed oscillations are
    proportional to $1/N^2$, while the frequency increases with $N$. }
  \label{fig:waves}
\end{figure}

Figure \ref{fig:waves} displays the fractional difference between the
early evolution of the Regge cosmology and the exact solution for the
axisymmetric case with $p_1 = p_2 = 2/3, p_3=-1/3$.  Although the
average magnitude of the discrepancy is small, there is a clear
oscillation superimposed upon a general trend.  The figure displays
data for three lattices consisting of $128\times4\times4$,
$256\times4\times4$ and $512\times4\times4$ vertices.

\begin{figure}[e]
  \begin{center}
    \epsfig{file=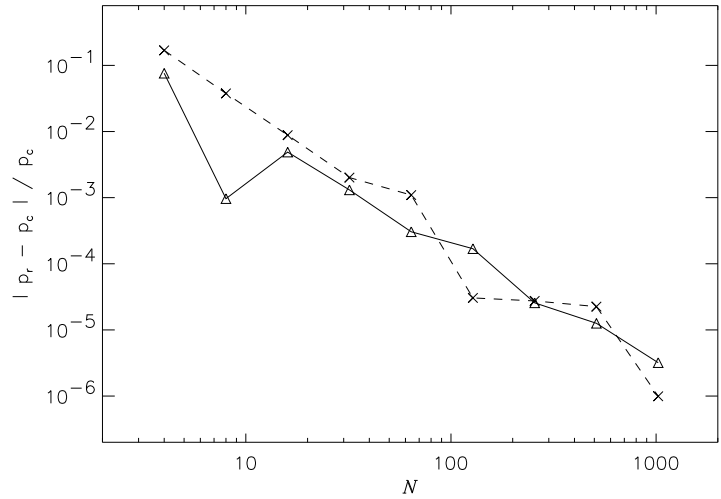,width=4.0in}
  \end{center}
  \caption{A least squares fit of the function $L_0 t^{p_r}$ (equation
    \ref{eqn:fit}) is performed on the edge lengths at different
    resolutions. We plot the fractional error in the fitted power
    $p_r$ as a function of the number of vertices. The fractional
    error in the fit along the $x$ and $y$-axes ($\triangle$-points;
    $p \approx 2/3$) are indistinguishable; the $z$-axis fit
    ($\times$-points; $p \approx -1/3$) follow the power-law
    behaviour.  The fractional error in the average Regge solution can
    thus be seen to vary as $1/N^2$. }
  \label{fig:fit}
\end{figure}

\begin{figure}[e]
  \begin{center}
    \epsfig{file=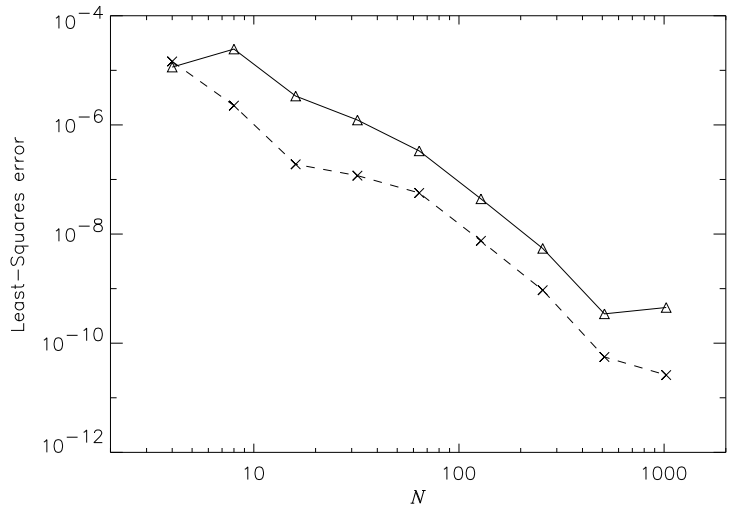,width=4.0in}
  \end{center}
  \caption{The error in the least squares fit of equation \eqn{fit} to
    the mean edge lengths is shown as a function of the number of
    vertices. The results for the $x$ and $y$-axes are again
    indistinguishable ($\triangle$), with the $z$-axis showing the
    same trend ($\times$).  This is a measure of both the accuracy of
    the functional form given in equation \eqn{fit} and the amplitude
    of the waves shown in figure \ref{fig:waves}.  We again find that
    the error measure converges to zero as $1/N^2$, indicating that
    the wave amplitude reduces as at least $1/N^2$. }
  \label{fig:noise}
\end{figure}

An estimate of the rate at which the Regge solution converges to the
continuum is given in figure \ref{fig:fit}.  After performing a least
squares fit of equation \eqn{fit} to the simplicial solution in the
region $t=[1,2]$, we plot the fractional difference between the fitted
power $p_r$ and the analytic value $p_c$ for each spatial axis.  It is
clear that this power law fit approaches the continuum solution as the
second power of the lattice spacing. Figure \ref{fig:noise} shows the
scaled error in the least squares fit, giving an indication of how
well the power law \eqn{fit} describes the Regge solution.  Again, we
find second order convergence.

The data represented in figure \ref{fig:noise} contains additional
information.  It is clear from figure \ref{fig:waves} that there is a
strong wave-like oscillation superimposed upon the general error
curve.  Fitting equation \eqn{fit} to such data will draw out the
averaged solution, while the error in the fit will reflect the
magnitude of any discrepancies from that average behaviour.  Thus we
see that both the oscillations and the underlying trend in figure
\ref{fig:waves} converge to zero as at least the second power of the
grid spacing.

\begin{figure}[e]
  \begin{center}
    \epsfig{file=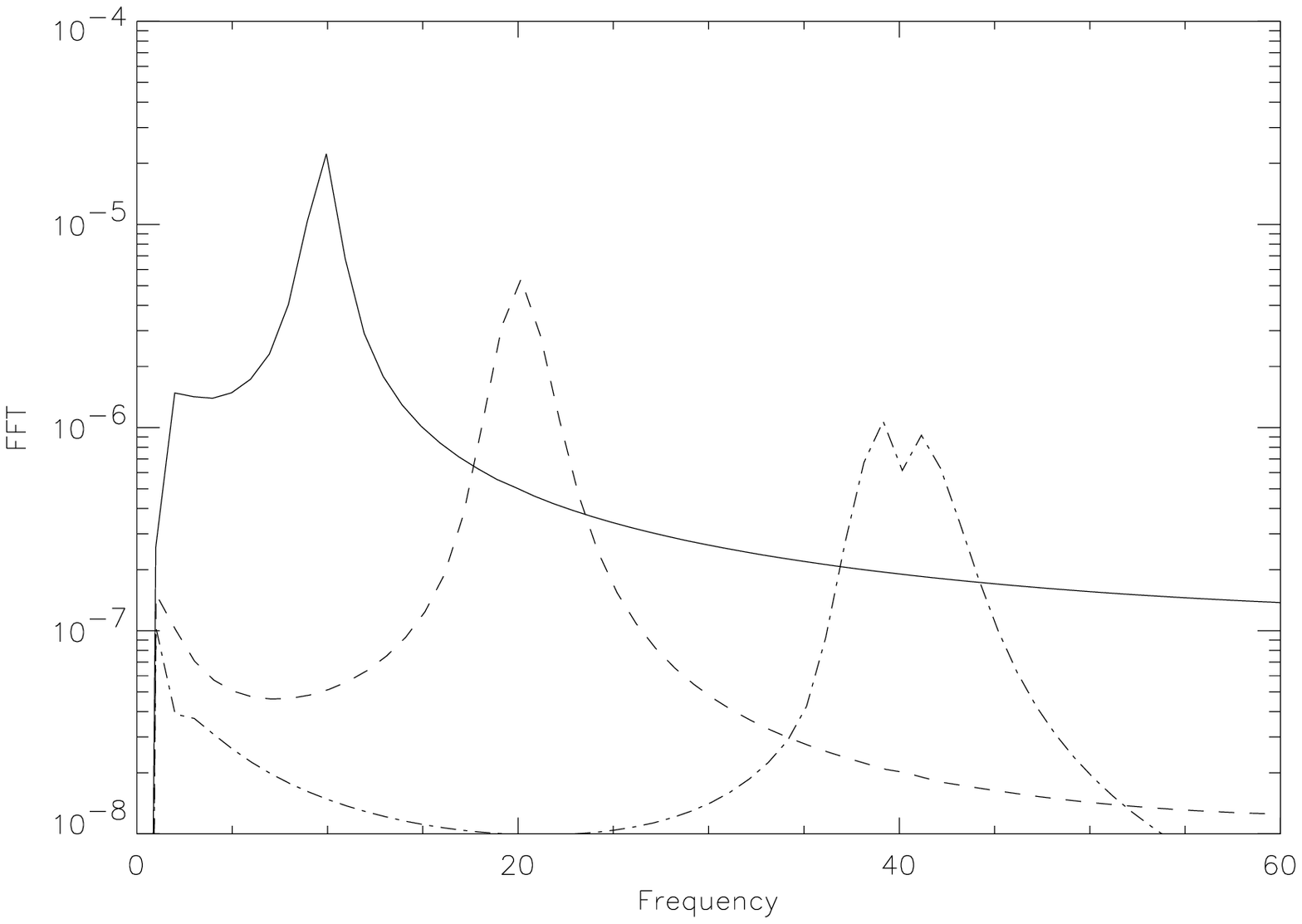,width=4.0in}
  \end{center}
  \caption{The Fourier transform of the data in figure 1 is shown for
    lattices containing of $128\times 4\times 4$ ($\full$), $256\times
    4\times 4$ ($\broken$), and $512\times 4\times 4$ ($\chain$)
    vertices.  The frequency of the wave-like oscillations apparent in
    figure \ref{fig:waves} are clearly proportional to $N$ (or
    $1/\delta$).  A quadratic drop-off in the power is also evident.}
   \label{fig:fft}
\end{figure}

In light of the discussion in the previous section, the frequency of
the oscillations evident in figure \ref{fig:waves} is also of
interest.  It is clear from the figure that the frequency increases
with the lattice resolution.  Figure \ref{fig:fft} shows this in a
more quantitative manner, displaying the Fourier transform of a small
segment of the fractional error.  It is apparent that the frequency of
the waves varies linearly with the number of vertices, or
alternatively, that the wavelength of the oscillations is proportional
to $1/N$.

Together, these $(3+1)$-dimensional convergence results show that the
numerical Regge solution is a second order accurate approximation to
the corresponding solution of the Einstein equations.  Moreover, the
leading order difference between the two solutions contains high
frequency, bounded oscillations.  We have shown that the frequency of
these waves is proportional to $N$ (or $1/\delta$), while their
magnitude reduces as $1/N^2$ ($\delta^2$).  This is precisely the
variation postulated by Brewin \cite{brewin95}, and strongly supports
his explanation of the apparent non-convergence of the Regge
equations.

Finally, we note that the Kasner solution displays only temporal
oscillations, with the spatial three-geometries remaining homogeneous
to high accuracy throughout the evolution.  This is likely a result of
the high degree of symmetry of the Kasner spacetime, in which the
constant time spatial hypersurfaces are flat. In a more general
setting we would expect to see similar high frequency, bounded
oscillations in both space and time.

\section{Conclusion}

We have carried out a rigorous convergence study of a particular
solution to the Regge equations, and shown that the solutions do
indeed converge to the corresponding Einstein solutions in the limit
of very fine discretisations.

Following a suggestion of Brewin we examined the behaviour of the
error terms, and found convincing evidence for the existence of high
frequency, low amplitude oscillations superimposed upon the continuum
solution.  Brewin previously suggested that waves of precisely this
form in the simplicial solution could explain the lack of convergence
observed in the residual of the Regge equations when they are
evaluated on the interpolated continuum solution.

Together these results suggest that solutions of the Regge equations
are generally second order accurate approximations to the
corresponding Einstein spacetimes, with the discrepancy between the
two solutions consisting of high frequency, low amplitude waves.
Although these waves prevent the residual of the Regge equations
converging to zero when evaluated on interpolated Einstein solutions
\cite{brewin95}, they do not affect the overall second order accuracy
of the simplicial solutions.

We expect that all generic Regge simulations in vacuum will produce
similar results.  That is, all simulations will contain high frequency
oscillations, linked to the inverse of the discretisation scale, which
reduce in magnitude as the second power of that scale.  These
oscillations do not appear to induce instabilities in the numerical
evolution of simplicial lattices. Indeed, previous studies have
indicated that the oscillations gradually decay as the evolution
proceeds \cite{gentle98}.  However, M.~Miller \cite{whybother} has
reported instabilities arising in linearised Regge calculus on
asymmetrical grids.

\section*{Acknowledgements}

We are grateful to Todd Lane for assistance with the preparation of
the figures, and the Australian Research Council for a Small Grant
to support this work.

\section*{References}

\gdef\journal#1, #2, #3, #4 {{#1}, {\bf #2}, #3 {(#4)}. }
\gdef\FP{\it{Found.~Phys.}}
\gdef\IJTP{\it{Int. J. Theor. Phys.}}
\gdef\IJMP{\it{Int. J. Mod. Phys.}}
\gdef\GRG{\it{Gen. Rel. Grav.}}
\gdef\PTP{\it{Prog. Theor. Phys.}}
\gdef\AP{\it{Ann. Phys.}}


\begin{thebibliography}{30}

\bibitem{regge61} T.~Regge,                      
  \journal \NC, 19, 558--71, 1961

\bibitem{gentle98} A.~P.~Gentle and W.~A.~Miller,
  \journal \CQG, 15, 389--405, 1998

\bibitem{gentle99} A.~P.~Gentle, 
  ``Regge Geometrodynamics'',
  {\it PhD Thesis}, Monash University (1999).
    
\bibitem{brewin95} L.~C.~Brewin,
  \journal \GRG, 32, 897--918, 2000
  
\bibitem{miller95} M.~A.~Miller,
  \journal \CQG, 12, 3037--51, 1995

\bibitem{private} R.~M.~Williams, W.~Miller, R.~D.~Sorkin, M.~Miller,
                 J.~W.~Barret, P.~A.~Tuckey
{\it private communications}.
    
\bibitem{sorkin75} R.~D.~Sorkin,
  \journal \JMP, 16, 2432--40, 1975

\bibitem{whybother} M.~A.~Miller,
  {\it PhD Thesis}, Syracuse University (1996).

\end{thebibliography}
\end{document}